# The comparative role of $^7$Li($n$,γ) reaction in primordial nucleosynthesis

N.A. Burkova[1], S.B. Dubovichenko[1,2,*], A.V. Dzhazairov-Kakhramanov[2,*] and S.Zh. Nurakhmetova[1]

[1] Al-Farabi Kazakh National University (KazNU), Almaty, Republic of Kazakhstan (RK)
[2] Fesenkov Astrophysical Institute, "NCSRT" ASC MDDIAI RK, Almaty, RK

**Abstract.** Within the framework of the modified potential cluster model with forbidden states and their classification according to Young diagrams, the possibility of describing experimental data on the total cross sections of the neutron radiative capture on $^7$Li is considered. It is shown that the model used and the methods for constructing potentials make it possible to correctly describe the behavior of experimental cross sections at energies of 5 meV ($5·10^{-3}$ eV) to 1 MeV($1·10^6$ eV), where experimental data are available. Based on the obtained total cross sections up to 5 MeV, the reaction rate was calculated and its analytical approximation was carried out. It was shown that the $^7$Li($n$,γ)$^8$Li reaction dominates at $T_9 < 0.2$ as against with the burning of $^7$Li in reactions $^7$Li($^3$H,$n$)$^9$Be and $^7$Li($^4$He,γ)$^{11}$B.

## 1 Introduction

As known, there are several chains of thermonuclear reactions of primordial nucleosynthesis, including the formation of unstable $^8$Li nucleus at a certain stage [1–4]

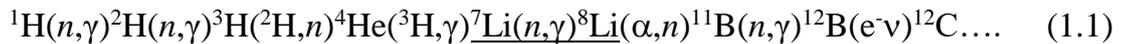

$$^1\text{H}(n,\gamma)^2\text{H}(n,\gamma)^3\text{H}(^2\text{H},n)^4\text{He}(^3\text{H},\gamma)\underline{^7\text{Li}(n,\gamma)^8\text{Li}}(\alpha,n)^{11}\text{B}(n,\gamma)^{12}\text{B}(e^-\nu)^{12}\text{C}\ldots \quad (1.1)$$

Here, a slightly different version of such a chain is possible, in which there is $^9$Be and characteristic for neutron-rich stars (see, for example, [5])

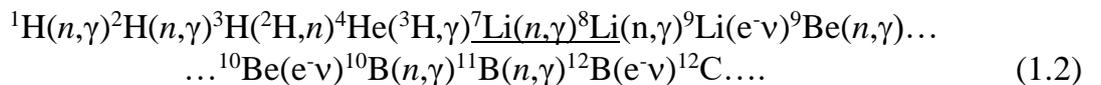

$$^1\text{H}(n,\gamma)^2\text{H}(n,\gamma)^3\text{H}(^2\text{H},n)^4\text{He}(^3\text{H},\gamma)\underline{^7\text{Li}(n,\gamma)^8\text{Li}}(n,\gamma)^9\text{Li}(e^-\nu)^9\text{Be}(n,\gamma)\ldots$$
$$\ldots^{10}\text{Be}(e^-\nu)^{10}\text{B}(n,\gamma)^{11}\text{B}(n,\gamma)^{12}\text{B}(e^-\nu)^{12}\text{C}\ldots \quad (1.2)$$

In some works, for example, [6–8], another option of such a chain was considered, similar to the previous boron process

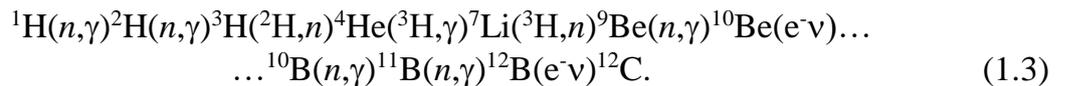

$$^1\text{H}(n,\gamma)^2\text{H}(n,\gamma)^3\text{H}(^2\text{H},n)^4\text{He}(^3\text{H},\gamma)^7\text{Li}(^3\text{H},n)^9\text{Be}(n,\gamma)^{10}\text{Be}(e^-\nu)\ldots$$
$$\ldots^{10}\text{B}(n,\gamma)^{11}\text{B}(n,\gamma)^{12}\text{B}(e^-\nu)^{12}\text{C}. \quad (1.3)$$

There is also a less likely chain option [1]

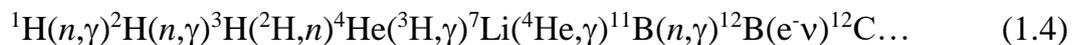

$$^1\text{H}(n,\gamma)^2\text{H}(n,\gamma)^3\text{H}(^2\text{H},n)^4\text{He}(^3\text{H},\gamma)^7\text{Li}(^4\text{He},\gamma)^{11}\text{B}(n,\gamma)^{12}\text{B}(e^-\nu)^{12}\text{C}\ldots \quad (1.4)$$

---

* Corresponding authors: albert-j@yandex.ru, dubovichenko@mail.ru

Several other scenarios of such and similar processes are given, for example, in [9,10]. As can be seen from these examples, the neutron capture process considered here on the $^7$Li nucleus is present in two versions of chains. As known, for calculations, for example, the abundance of chemical elements in the Universe, the rates of all the above reactions are required [9–12]. However, as it will be seen later, not all of them are known to date with the required degree of accuracy. Different options of calculations this reaction rate may vary by 1–1.5 orders of magnitude. Therefore, calculations of such rates within the framework of microscopically reasonable nuclear models, such as the modified potential cluster model (MPCM) used by us [13–15], can solve the existing problem of insufficient accuracy in determining the rates of thermonuclear reactions (see, for example, [16]).

In this paper, we will consider the possibility of describing the total cross sections of the reaction of the neutron radiative capture on $^7$Li at energies of 5 meV to 1 MeV, where the experimental data are known within the framework of the MPCM. This model takes into account the supermultiplet symmetry of the wave function (WF) with the separation of orbital states according to Young diagrams. This approach allows us to analyze the structure of intercluster interactions, to determine the presence and position of the allowed (AS) and forbidden states (FS) in intercluster potentials.

Within the framework of the MPCM used, the potentials of scattering processes are usually constructed on the basis of a description of the elastic scattering phase shifts, taking into account their resonance behavior, which are extracted in the phase shift analysis from experimental differential cross sections. Interactions of bound states (BS) are determined from the additional requirement that they reproduce some characteristics of the ground state (GS) of the nucleus, including the asymptotic constant (AC) in a particular channel. It is assumed that this BS is mainly caused by the cluster channel consisting of the initial particles, which participate in the reaction. As a result, the model does not have free parameters and allows one to describe correctly the total cross sections of the radiative capture.

Earlier, in the framework of the MPCM, we considered almost 40 such reactions [13–15,17,18] (see references in these works). In addition, similar results can be found, for example, in [19] for some processes of radiative capture. We have already considered the neutron capture reaction on $^7$Li in [20], but the reaction rate was not calculated. In this article, based on the obtained total cross sections of 1 meV to 5 MeV, the reaction rate of [21] the $^7$Li($n,\gamma$)$^8$Li capture will be calculated at temperatures of 0.01 to 10 $T_9$.

## 2 Calculation methods

Total cross sections of the radiative capture σ($NJ,J_f$) for *EJ* and *MJ* transitions in the cluster model are given, for example, in [13–15,22] and have the form

$$\sigma(NJ, J_f) = \frac{8\pi K e^2}{\hbar^2 q^3} \frac{\mu}{(2S_1+1)(2S_2+1)} \frac{J+1}{J[(2J+1)!!]^2} A_J^2(NJ,K) \cdot$$
$$\cdot \sum_{L_i,J_i} P_J^2(NJ,J_f,J_i) I_J^2(J_f,J_i), \qquad (2.1)$$

where $\sigma$ is the total cross section of the radiative capture process, $\mu$ is the reduced mass of particles of the initial channel, $q$ is the wave number of the initial channel particles $q^2 = 2\mu E/\hbar^2$ and $E$ is the energy of particles in the initial channel, $S_1$, $S_2$ are spins of initial particles, $K = E_\gamma/\hbar c$ and $J$ – wave number and momentum γ-quantum of the output channel, $N$ is $E$ or $M$ transitions of the $J$ multipolarity from the initial $J_i$ to the final $J_f$ state of the nucleus under consideration.

For electric orbital $EJ(L)$ transitions ($S_i = S_f = S$) the values of $P_J$ and $A_J$ have the form

$$P_J^2(EJ, J_f, J_i) = \delta_{S_i S_f} \left[(2J+1)(2L_i+1)(2J_i+1)(2J_f+1)\right](L_i 0 J 0 | L_f 0)^2 \begin{Bmatrix} L_i & S & J_i \\ J_f & J & L_f \end{Bmatrix}^2$$

$$A_J(EJ, K) = K^J \mu^J \left(\frac{Z_1}{m_1^J} + (-1)^J \frac{Z_2}{m_2^J}\right), \qquad I_J(J_f, J_i) = \langle \chi_f | r^J | \chi_i \rangle. \qquad (2.2)$$

Here $L_f$, $L_i$, $J_f$, $J_i$ are momenta of particles of the initial and final channels, $m_1$, $m_2$, $Z_1$, $Z_2$ are masses and charges of particles of the initial channel, $I_J$ is the integral over wave functions of the initial $\chi_i$ and final $\chi_f$ states, as functions of the relative motion of clusters with intercluster distance $r$.

For the spin part of the magnetic process $MJ(S)$ in the using model, the expression ($S_i = S_f = S$, $L_i = L_f = L$) was obtained [13–16]

$$P_J^2(MJ, J_f, J_i) = \delta_{S_i S_f} \left[S(S+1)(2S+1)(2J_i+1)(2L_i+1)(2M+1)(2J+1)(2J_f+1)\right] \cdot$$

$$\cdot (L_i 0 M 0 | L_f 0)^2 \begin{Bmatrix} L_i & M & L_f \\ S & 1 & S \\ J_i & J & J_f \end{Bmatrix}^2,$$

where $M = J - 1$ and

$$A_J(MJ, K) = \frac{\hbar K}{m_0 c} K^{J-1} \sqrt{J(2J+1)} \left[\mu_1 \left(\frac{m_2}{m}\right)^J + (-1)^J \mu_2 \left(\frac{m_1}{m}\right)^J\right], \qquad (2.3)$$

$$I_J(J_f, J_i) = \langle \chi_f | r^{J-1} | \chi_i \rangle.$$

Here $m$ is the mass of nucleus, $\mu_1$, $\mu_2$ are magnetic momenta of clusters, other designations as in the previous expression. We get to these values for the magnetic momenta of the neutron and $^7$Li: $\mu_n = -1.91304272\mu_0$ [23] and $\mu(^7\text{Li}) = 3.256427\mu_0$ [24]. In all our calculations, the exact values of the particle masses were set ($m_n = 1.00866491597$ amu and $m_{Li} = 7.01600455$ amu) [23,24], and constant $\hbar^2/m_0$, assumed equal to 41.4686 MeV fm$^2$.

If there are known total cross sections of the reaction, in this case the radiative capture process, then it is possible to determine reaction rate of this reaction as

$$N_A \langle \sigma v \rangle = 3.7313 \cdot 10^4 \mu^{-1/2} T_9^{-3/2} \int_0^\infty \sigma(E) E \exp(-11.605 E / T_9) dE, \qquad (2.4)$$

where $E$ – energy is given in MeV, cross section $\sigma(E)$ measures in μb, $\mu$ is the reduced mass in amu, $T_9$ is the temperature in $10^9$ K.

## 3 Structure of n$^7$Li states

In our previous works [13–15,20] it was shown that if the Young diagram{43} [25,26] is used for the $^7$Li nucleus, then for the n$^7$Li, p$^7$Li or p$^7$Be, n$^7$Be systems we get {1} + {43} = {53} + {44} + {431}. Therefore, FSs are present in the $^3P$ waves with the {53} diagram and in the $^3S_1$ wave for {44}. The allowed $^3P$ states have the {431} diagram. Thus, n$^7$Li potentials in the triplet spin state must have a forbidden bound $^3S_1$ state with the {44} diagram for scattering processes, which will be considered later, and forbidden and allowed bound levels in $^3P$ waves with Young diagrams {53} and {431 }, the last of which corresponds to the $^3P_2$ ground bound state of $^8$Li in the n$^7$Li channel. Note that the forbidden diagram {53} can also correspond to the momentum $L = 3$ [13–15], i.e., the FS can be present in the $F$ wave.

All these results relate to the triplet spin n$^7$Li state, and for spin $S = 2$ the allowed symmetries, and, consequently, the bound allowed levels in the n$^7$Li system, are absent for any values of the orbital momentum $L$ [27]. Thus, the potential of the $^5S_2$ scattering wave has a bound FS with the {44} diagram, and in the $^5P$ waves of the scattering process, the potential contains only FS with the {53} and {431} diagrams, the last of which can be in the continuous spectrum, and the potential has only one bound FS with the {53} diagram.

```
  4.650, (1⁺1)
  ─────────────

  3.210, (1⁺1)
  ─────────────

  2.255(3), (3⁺1)
  ─────────────
                        n⁷Li
  0.9808(1), (1⁺1)   2.03229
  ─────────────

  ─────────────
  ⁸Li, (2⁺1)
```

**Figure 1.** Spectrum of $^8$Li [29,30] nucleus levels in MeV

Generally, the scattering potentials are constructed based on the results of the phase shift analysis of the experimental cross sections for elastic scattering. However, we were unable to find data on the phase shift analysis of the n$^7$Li elastic scattering at astrophysical energies [28]. Therefore, here we will construct the potentials of scattering processes in the n$^7$Li system by analogy with p$^7$Li scattering [13–15] using data on the spectrum of $^8$Li levels [29,30], which is shown in figure 1.

The bound state of the n$^7$Li system with $J^\pi T = 2^+1$, corresponding to the GS of $^8$Li, can form at $S = 1$ and 2 with the orbital momentum $L = 1$ and, in the general case, is a mixture of $^3P_2$ and $^5P_2$ states. Despite absence of the AS at $S = 2$, as follows from the above classification, the presence of an impurity of the $^5P_2$ wave in the WF BS should be accepted. This is necessary for further consideration of the $M1$ transition from the first $^5P_3$ resonance state in the n$^7$Li scattering with $J^\pi T = 3^+1$ at 2.255 MeV or 0.22 MeV in center-of-mass system (c.m.) relative to the threshold at a width of 35(5) keV on the $^5P_2$ WF component of the GS of $^8$Li. Such resonance of the $^5P_3$ state, if we consider the orbital momenta with the lowest possible values, can be formed only with the total spin of the n$^7$Li system $S = 2$.

The first excited state (FES) with $J^\pi T = 1^+1$ is the $^{3+5}P_1$ level in the n$^7$Li channel and turns out to be bounded at an energy of 0.9808 MeV relative to the GS of $^8$Li or –1.05149 MeV relative to the threshold of the n$^7$Li channel [29,30]. Furthermore, we consider the $E1$ transitions to this level from triplet and quintet $S$ scattering waves. Therefore, all further results will relate to the reactions $^7$Li(n,γ$_0$)$^8$Li and $^7$Li(n,γ$_1$)$^8$Li and the sum of their cross sections. Furthermore, based on the data of [29,30], we assume that the $^3S_1$ and $^5S_2$ phase shifts in the range up to 3.0 MeV are close to zero and have a smoothly decreasing shape. This is confirmed by the absence of resonance levels with negative parity at such energies in the spectrum of $^8$Li, shown in figure 1.

The second and third resonance states with $J^\pi T = 1^+1$ lies at energies above the threshold of 1.18 MeV and 2.03 MeV with widths of ~1 MeV and 0.1 MeV [29]. For these states, their momenta [20] and widths [30] were not known before. They can also be considered as $^{3+5}P_1$ or $^5F_1$ scattering states of the $n^7$Li system. In addition to them, some resonances with the widths and momenta are still not known [29], so we will not consider them.

## 4 Interaction potentials

To perform calculations of photonuclear processes in the cluster systems under consideration, the nuclear part of the intercluster interaction potential is represented in the form [13–18]

$$V(r) = -V_0 \exp(-\alpha r^2) \quad (4.1)$$

with the Gaussian attracting part $V_0$.

The asymptotic constant of the GS $^8$Li potential in the $n^7$Li channel is calculated using the WF asymptotics in the form of the Whittaker function [31]

$$\chi_L(R) = \sqrt{2k} C_W W_{-\eta, L+1/2}(2kR), \quad (4.2)$$

where $\chi$ is the numerical wave function of the bound state obtained from the solution of the radial Schrödinger equation and normalized to unity, $W$ is the Whittaker function of the bound state, which determines the asymptotic behavior of the WF at $r = R$ and is a solution of the same equation without nuclear potential, i.e., the solution at large distances $R$, $k$ is the wave number dependents from the channel binding energy, $\eta$ is the Coulomb parameter, which in this case eguals zero, $L$ is the orbital momentum of the bound state of clusters.

The most complete versions of potentials and calculations with them were considered by us in [20,32], but the reaction rate was not calculated for them. Here we briefly dwell on only one version of the potentials that corresponds to the above classification of Young diagrams. The parameters of such interactions for different partial waves are listed in Table 1.

**Table 1.** Parameters of potentials for $n^7$Li interactions in different partial waves.

| No. | $^{2S+1}L_J$ | $V_0$, MeV | $\alpha$, fm$^{-2}$ | $E_R$ (c.m.), keV | $\Gamma_R$ (c.m.), keV |
|---|---|---|---|---|---|
| 1. | Nonresonance $^3S_1$, $^5S_2$ scattering waves | 50.5 | 0.15 | --- | --- |
| 2. | Nonresonance $^3P_0$, $^{3+5}P_2$ scattering waves | 132.0 | 0.1 | --- | --- |
| 3. | Resonance $^5P_3$ scattering wave | 2059.75 | 2.5 | 223(1) | 35(1) |
| 4. | Resonance $^{3+5}P_1$ scattering wave | 482.16 | 0.6 | 1180(1) | 1027(10) |
| 5. | Resonance $^5F_1$ scattering wave | 251.04 | 0.13 | 2030(1) | 106(1) |
| 6. | Bound $^{3+5}P_2$ GS | 429.383779 | 0.5 | $E_b = -2.0322900$ | $C_W = 0.78(1)$ |
| 7. | Bound $^{3+5}P_1$ FES | 422.126824 | 0.5 | $E_b = -1.051490$ | $C_W = 0.59(1)$ |

The zero phase shifts of the $S$ scattering in any spin channel can be obtained with potential No. 1 from Table 1, which has one bound FS for the given classification option with the {44} diagram. Since the potential has FS, its phase shift starts from 180° [19], and the phase shift shape is shown in figure 2 by the dotted-dashed curve. For nonresonance $^3P_0$, $^{3+5}P_2$ scattering waves with one bound FS, on conditions that the BS with {431} is in the continuous spectrum, we use the parameters No. 2 from Table 1, which lead to scattering phase shifts in the range of 180(2)° at energies up to 3 MeV.

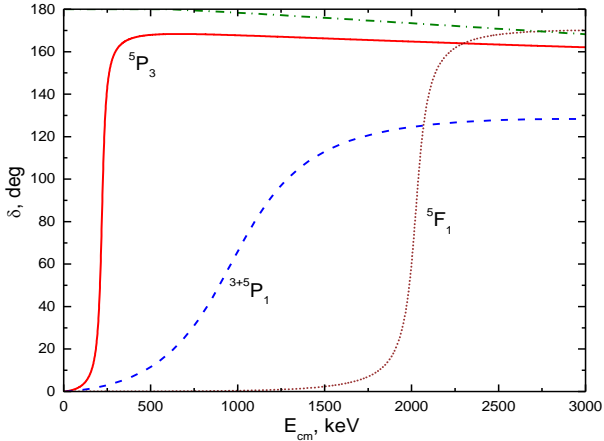

Figure 2. Elastic $n^7$Li scattering phase shifts.

The resonance $^5P_3$ phase shift of the elastic $n^7$Li scattering can be described by a Gaussian potential with parameters No. 3 from Table 1. Such a potential has one bound FS with the {53} diagram and also corresponds to the above classification, provided that the BS with {431} is in the continuous spectrum. The results of the phase shift calculation are shown in figure 2 by the solid curve – the resonance is at the energy of 223(1) keV centre-of-mass system (c.m.). The width of the $^5P_3$ resonance is equal to 35(1) keV (c.m.) at experimental values of 35(5) keV or 33(6) keV according to various data from reviews [29,30]. The resonance width is determined according to the expression $\Gamma_{c.m.} = 2(d\delta/dE_{c.m.})^{-1}$.

In this paper, we expand the energy range under consideration to 5 MeV and consider the second and third resonances shown in figure 1. The parameters for the potential of the second resonance state (No. 4 in Table 1) were obtained. They allow one to correctly describe its energy and reproduce well the order of the width, and the phase shift of such a potential is shown in figure 2 by the dashed curve.

We were not able to obtain the potential parameters of the third resonance supposing that it is the $P$ wave potential, so as it was able to correctly describe not only the energy, but also the width. Even a potential width parameter of 10 fm$^{-2}$ made it possible to obtain a resonance width of about 500 keV at an experimental value of 100 keV. Therefore, we assumed that such resonance is in the $^5F_1$ wave and immediately obtained the parameters given in Table 1 under No. 5. Such a potential with the FS allows one to obtain correctly both the resonance energy and its width, and its phase shift is shown in figure 2 by the frequent dotted curve. The phase shifts of $P$ and $F$ waves Nos. 3, 4 and 5 from Table 1 are shown in figure 2 from zero degrees, i.e., in a more familiar form, although they should start at 180°, since they have FS.

For the $^{3+5}P_2$ GS potential of the $n^7$Li system, which corresponds to the state of $^8$Li in the considering cluster channel, it is possible to use parameters No. 6 [20,32]. Besides to the allowed BS for the {431} diagram corresponding to the GS of $^8$Li, such $^{3+5}P_2$ potential has the FS for the {53} diagram in full accordance with the classification of orbital states given above. The binding energy of –2.03229 MeV was obtained with such a potential, which completely coincides with the experimental value [30], charge and mass radii are equal to 2.38 fm and 2.45 fm, respectively. We use the value equals to the proton radius of 0.8775(51) fm [23] as the neutron mass radius. Apparently, the root-mean-square (rms) charge radius of $^8$Li should not appreciably exceed the radius of $^7$Li, which equals 2.35(10) fm [33]. Therefore, the rms radius value obtained above for the $n^7$Li channel in the GS of $^8$Li is quite reasonable. The asymptotic constant for such GS potential turned out to be $C_w = 0.78(1)$. The AC error is determined by its averaging over the range 5–30 fm, where it was relatively stable.

Parameters No. 7 were obtained for the FES potential. Here, the allowed BS for {431} diagram corresponds to the FES of $^8$Li at 0.9808 MeV. In addition, this $^{3+5}P_1$ potential has the FS for {53} in full accordance with the above mentioned classification of orbital states. The binding energy of –1.05149 MeV, which completely coincides with the experimental value [29,30], the charge radius 2.39 fm, and the mass radius 2.52 fm were obtained with this potential. The asymptotic constant of this potential turned out to be $C_w = 0.59(1)$. The AC error is determined by its averaging over the range 4–25 fm, where the asymptotic constant remains stable.

To compare our results on AC with the available data, we note that, for example, in [34] and some other works, a slightly different definition of AC was used, which has the dimension

$$\chi_L(R) = C \cdot W_{-\eta L+1/2}(2kR). \qquad (4.3)$$

This AC is related to our $C_w$ through the value $\sqrt{2k}$. The asymptotic normalization coefficient (ANC) $A_{NC}$, which is determined on the basis of experimental data, is associated with the dimensional AC through the spectroscopic factor $S_f$

$$A_{NC}^2 = S_f \cdot C^2. \qquad (4.4)$$

Here are some results for the ANC of the $n^7$Li system. So in [34], from the analysis of experimental data, the value $C = 0.657(34)$ fm$^{-1/2}$ was obtained. After recalculation to $\sqrt{2k} = 0.767$ a dimensionless quantity at and with a unit spectrum factor $S_f$, the dimensionless AC turned out to be $C_w = 0.857(44)$. However, the results for $S_f$ of the system under consideration were presented in [35], which can be represented in the form 0.9(2). If such a spectroscopic factor is taken into account, then for a dimensionless AC, the value $C_w = 0.92(15)$ is obtained. In one of the latest works [36], $C^2(p_{3/2}) = 0.43(11)$ and $C^2(p_{1/2}) = 0.056(16)$ were obtained, which for $S_f = 1$ gives $C_w = 0.90(12)$. If we use $S_f = 0.9(2)$ from [35], then the interval for $C_w$ expands and becomes equal to 1.08(38). Thus, the AC values obtained in these studies are in the range 0.70–1.46.

The value $C = 0.082(9)$ fm$^{-1/2}$ is given for the FES of $^8$Li in [34], and the spectroscopic factor $S_f = 0.4(1)$ is given in [35]. Using these two quantities for the dimensionless AC $C_w$, we obtain 0.71(1). Note that the paper [37] provides a summary table of ANC results obtained in various studies, and it is precisely the two articles considered above [34,36] that allow us to obtain the closest results. It can be seen from this that the calculated AC of the GS and FES potentials is located at the lower boundary of these intervals. Therefore, here we will use the GS and FES potentials, which were proposed earlier in our works [20,32].

# 5  Reaction rate of the radiative $^7$Li(n,γ)$^8$Li capture

Before our works [20,32], the $E$1 transition in the $n^7$Li system at the radiative capture was considered in [38–40], where the possibility of the correct description of the total cross sections in the nonresonance energy range was shown. As for the resonance at 0.22 MeV, in [41], on the basis of model-independent calculation methods, results were obtained with an acceptable description of the resonance cross sections of the radiative capture.

When considering electromagnetic processes in the $^7$Li($n$,γ)$^8$Li reaction, we will take into account the $E$1 transition from the nonresonance $S$ scattering waves to the $^{3+5}P_2$ GS WF. Additionally, we will take into account the $M$1 transition from the resonance $^5P_3$ scattering wave to the quintet $^5P_2$ part of the GS WF of this nucleus. Since it is not possible directly to extract the $^5P_2$ and $^3P_2$ parts of the GS WF in this model, we will use the spin-mixed function that is obtained with the proposed BS potential for the calculations performed.

**Table 2.** Coefficients $P^2$ in cross sections (2.2) and (2.3) for the considered transitions.

| No. | $^{(2S+1)}L_J)_i$ | Transition | $^{(2S+1)}L_J)_f$ | $P^2$ |
|---|---|---|---|---|
| 1. | $^3S_1$ | $E$1 | $^3P_2$ | 5 |
| 2. | $^5S_2$ | $E$1 | $^5P_2$ | 5 |
| 3. | $^3P_1$ | $M$1 | $^3P_2$ | 5/2 |
| 4. | $^5P_1$ | $M$1 | $^5P_2$ | 9/2 |
| 5. | $^3P_2$ | $M$1 | $^3P_2$ | 15/2 |
| 6. | $^5P_2$ | $M$1 | $^5P_2$ | 125/6 |
| 7. | $^5P_3$ | $M$1 | $^5P_2$ | 14/3 |
| 8. | $^5F_1$ | $E$2 | $^5P_2$ | 3/5 |
| 9. | $^3S_1$ | $E$1 | $^3P_1$ | 3 |
| 10. | $^5S_2$ | $E$1 | $^5P_1$ | 5 |
| 11. | $^3P_1$ | $M$1 | $^3P_1$ | 3/2 |
| 12. | $^5P_1$ | $M$1 | $^5P_1$ | 27/2 |
| 13. | $^3P_2$ | $M$1 | $^3P_1$ | 5/2 |
| 14. | $^5P_2$ | $M$1 | $^5P_1$ | 9/2 |
| 15. | $^5F_1$ | $E$2 | $^5P_1$ | 81/25 |

Then, the total cross section of the capture process, taking into account all electromagnetic transitions considered here to the GS of $^8$Li for the neutron capture on $^7$Li, can be represented as

$$\sigma_0(E1+M1) = \sigma(E1,^3S_1 \to {}^3P_2) + \sigma(E1,^5S_2 \to {}^5P_2) + \sigma(M1,^5P_3 \to {}^5P_2) + \\ + [\sigma(M1,^3P_2 \to {}^3P_2) + \sigma(M1,^5P_2 \to {}^5P_2)]/2 + \\ + [\sigma(M1,^3P_1 \to {}^3P_2) + \sigma(M1,^5P_1 \to {}^5P_2)]/2 \quad . \quad (5.1)$$

Since the $P_1$ and $P_2$ scattering waves and the GS WF are mixed by spins, their averaging is used, as was already done in our previous works [13,14,42]. We note right away that the $E$2 transition from the $^3F_1$ wave with potential No. 5 does not make any noticeable contribution to the total cross sections at the capture to the GS or FES.

For transitions to the first excited $^{3+5}P_1$ state, the $E1$ process from both $^{3+5}S$ scattering waves and $M1$ processes from $P$ waves were taken into account

$$\sigma_1(E1+M1) = \sigma(E1,^3S_1 \to ^3P_1) + \sigma(E1,^5S_2 \to ^5P_1) + $$
$$+ \left[\sigma(M1,^3P_1 \to ^3P_1) + \sigma(M1,^5P_1 \to ^5P_1)\right]/2 + \qquad (5.2)$$
$$+ \left[\sigma(M1,^3P_2 \to ^3P_1) + \sigma(M1,^5P_2 \to ^5P_1)\right]/2$$

The results of the calculations for all these transitions were compared with experimental measurements of the total cross sections of the capture reaction in the energy range from 5 meV to 1.0 MeV, performed in [1,43–49]. Coefficients $P^2$ for the noted transitions are listed in Table 2.

Figure 3a shows the results of calculations for the summarized total cross section over all transitions of the neutron radiative capture on $^7$Li to the GS at energies of 5 keV to 1.5 MeV by the dotted-dashed curve. These results were obtained for the GS potential No. 6, $S$ scattering waves in the triplet and quintet states with parameters No. 1, and the potential of the resonance $^5P_3$ scattering wave with parameters No. 3. The dashed curve shows the cross section corresponding to the sum of the $E1$ transitions from $^3S_1$ and $^5S_2$ waves to the GS, and the dotted curve shows the cross section of the $M1$ transition between the $^5P_3$ scattering state and the GS of $^8$Li.

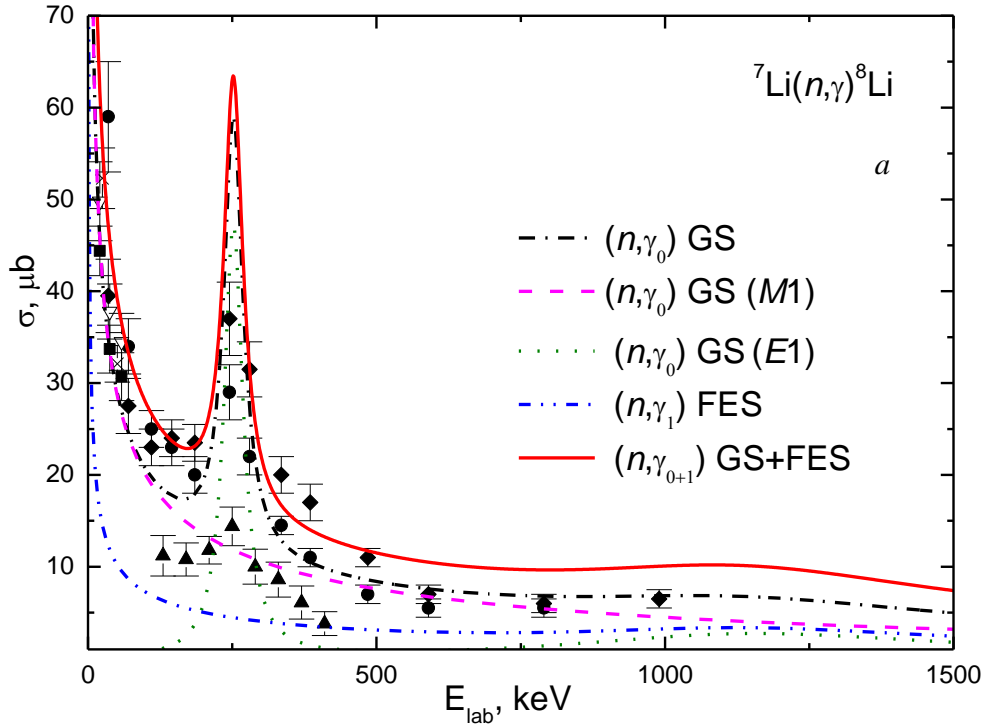

**Figure 3a**. Total reaction cross sections for the neutron radiative capture on $^7$Li at low energies. Experimental data: ● and ♦ – from work [43], ■ – [44] for capture to the GS and ∇ – summarized total sections for capture to the GS and FES, ▲ – [45], × – [46].

This cross section takes into account transitions from other $P$ waves. The dotted-dotted-dashed curve shows the results for capture to the FES. For the FES, the potential No. 7 was used and the same potentials for other scattering waves from Table 1 were used. The solid curve shows the results for total cross sections, taking into account all transitions to the GS and FES. The first resonance is observed at 252 keV laboratory system (l.s.) and has a value of 63 μb. It can be seen from figure 3a that taking into account the resonance at 1.18 MeV with potential No. 4 leads to a noticeable increase in the total cross sections in this range.

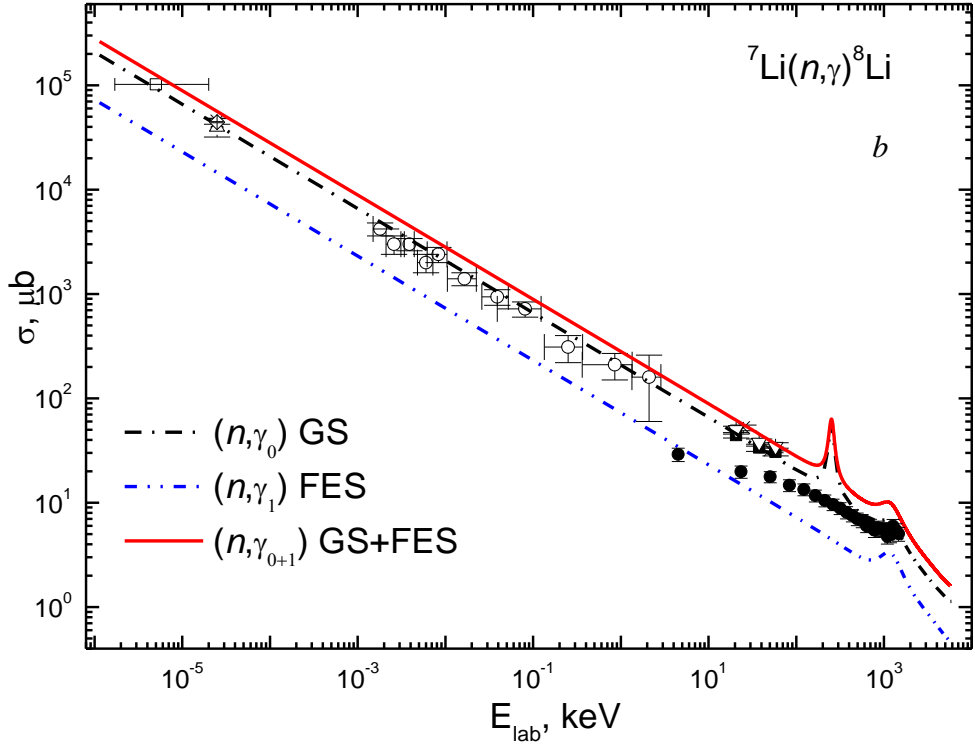

**Figure 3b.** Total reaction cross sections for the neutron radiative capture on $^7$Li at thermal energies. Experimental data: ■ – [44] for capture to the GS and ∇ – summarized total sections for the capture to the GS and FES, ○ – [47], □ – [1], Δ – [43], * – results of work [48] and other given data in it, × – [46], ◊ – last data for thermal cross section [49], measurements [50] are shown by dots ●. Curves – calculation results for different transitions with potentials given in the text.

Figure 3b shows to some detail the shape and magnitude of the total calculated capture cross sections at energies of 1 meV to 5 MeV – the designation of the curves, as in the previous figure. Here, the second resonance at 1.18 MeV is clearly visible, which is present upon capture to the GS and FES. It should be noted that in [47], the results of which are shown by open circles in figure 3b, the cross sections were measured only for the capture to the GS of $^8$Li. In [44], measurements were carried out for capture both to the GS (black squares) and for total cross sections taking into account transitions to the GS and FES (inverted open triangles with the base up). In addition, the data in [43] have a maximum at 245 keV (l.s.) with a value of 37 μb, and the resonance is at the experimental energy of 255 keV (l.s.) and there are no available data for it. The new results from [50] are in poor agreement with previous measurements and absolutely do not contain resonance at 255 keV.

Based on the results obtained, we can assume that the use of the GS and FES potentials of $^8$Li in the $n^7$Li channel with one FS and the corresponding scattering potentials leads to a reasonable description of the available experimental data for the total radiative capture cross sections in the entire energy range considered, whose boundaries differ by almost ten orders of magnitude.

Since at energies of $10^{-6}$ keV (1 meV) and up to about 0.1 keV, the calculated cross section shown in figure 3b by the solid curve is almost a straight line, it can be approximated by a simple function of the form

$$\sigma(\mu b) = \frac{A}{\sqrt{E_n(\text{keV})}}. \qquad (5.3)$$

The value of the given constant $A = 263.192$ µb·keV$^{1/2}$ was determined from a single point in the cross sections with a minimum energy of $10^{-6}$ keV. This value is slightly different from our previously obtained 265.738 µb keV$^{1/2}$ [14,18], but this difference does not exceed 1%, which can be explained by an increase in the accuracy of the present calculations – earlier, we used finite-difference methods to calculate the WF, and in the present work, a more accurate Numerov method is used.

We can consider the module of the relative deviation of the calculated theoretical total cross section and approximation of this cross section by function (5.3) in the energy range of $10^{-6}$ to 100 eV:

$$M(E) = \left| [\sigma_{ap}(E) - \sigma_{theor}(E)] / \sigma_{theor}(E) \right|, \qquad (5.4)$$

which at energies below 100 eV does not exceed 1%.

Furthermore, the reaction rate under consideration was calculated, the shape of which is shown in figure 4 – the solid, dotted-dashed and dotted-dotted-dashed curves correspond to the same curves in figure 3. Figure 4 shows in detail the obtained reaction rate and its comparison with some other reports.

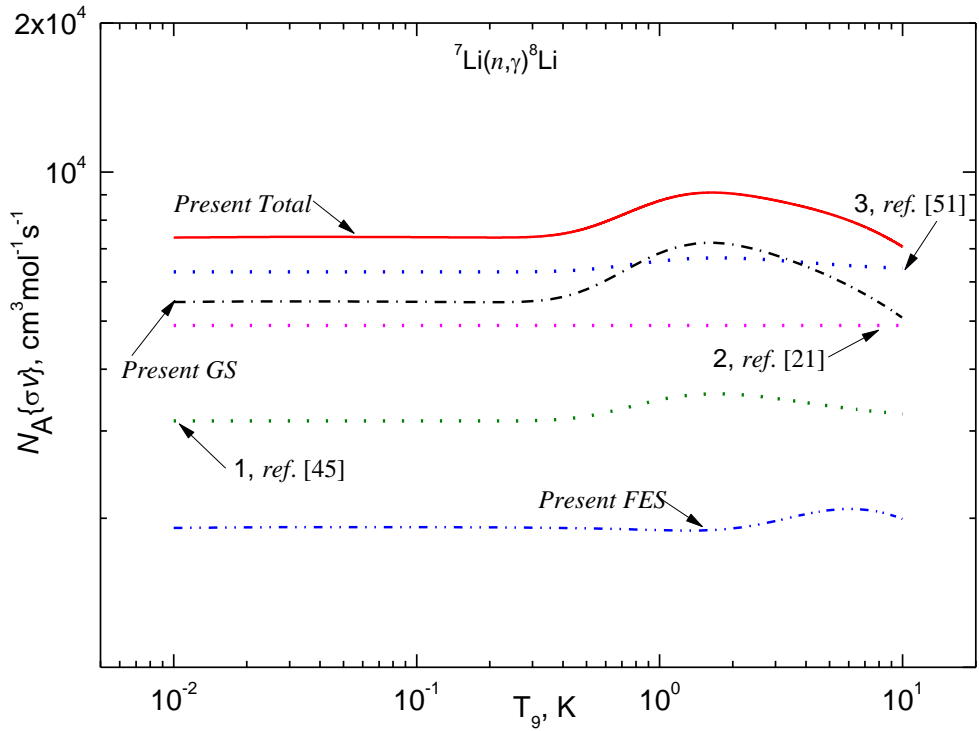

**Figure 4.** The comparative results of $^7$Li(n,γ)$^8$Li reaction rates. Curves are explained in the text.

In figure 4, we compare the obtained reaction rate with the results of [45] (dotted curve 1) for the neutron capture on $^7$Li. Dotted curve 2 shows the results from [21]. The dotted curve 3 is the corresponding parametrization from [51].

We used the expression [21] to approximate the obtained reaction rate of $^7$Li$(n,\gamma)^8$Li, which is shown in figures 4,5 by the solid curve

$$N_A \langle \sigma v \rangle = a_1 + \frac{a_2}{T_9^{5/2}} \exp\left(-\frac{a_3}{T_9^{1/2}}\right) + a_4 T_9^{a_5}, \qquad (5.5)$$

and the approximation parameters are given in Table 2.

Table 2. The approximation parameters of the $^7$Li$(n,\gamma)^8$Li capture reaction rate (5.5)

| $a_1$ | $a_2$ | $a_3$ | $a_4$ | $a_5$ |
|---|---|---|---|---|
| 0.7306979E+04 | 0.1116921E+07 | -0.6682375E+01 | -0.9260274E+00 | 0.2871425E+01 |

With such parameters, a value of $\chi^2 = 0.005$ is obtained with 5% errors of the calculated reaction rate, and the results of such parameterization practically merges with the solid curve.

In figure 5, dashed curve 2 – the alpha-particle capture reaction rate of $^7$Li$(^4$He$,\gamma)^{11}$B [52]. The dashed curve 3 shows the $^7$Li$(^3$He$,p_0)^9$Be reaction rate [53]. Dotted curve 4 illuminates the approximation of the $^7$Li$(^3$H$,n)^9$Be reaction rate [54]. The dotted curve 5 in figure 5 refers to the $^7$Li$(^3$H$,n)^9$Be reaction rate from [7], obtained up to 3 $T_9$ and practically does not differ from the results of [8]. The $^7$Li$(^3$H$,n)^9$Be reaction rate to all bound states is shown by the frequent dashed curve 6, and the frequent dotted curve 7 is for the reaction rate only to the GS of $^9$Be from [8], in this work, the $\langle \sigma v \rangle$ is given starting from 0.1 $T_9$.

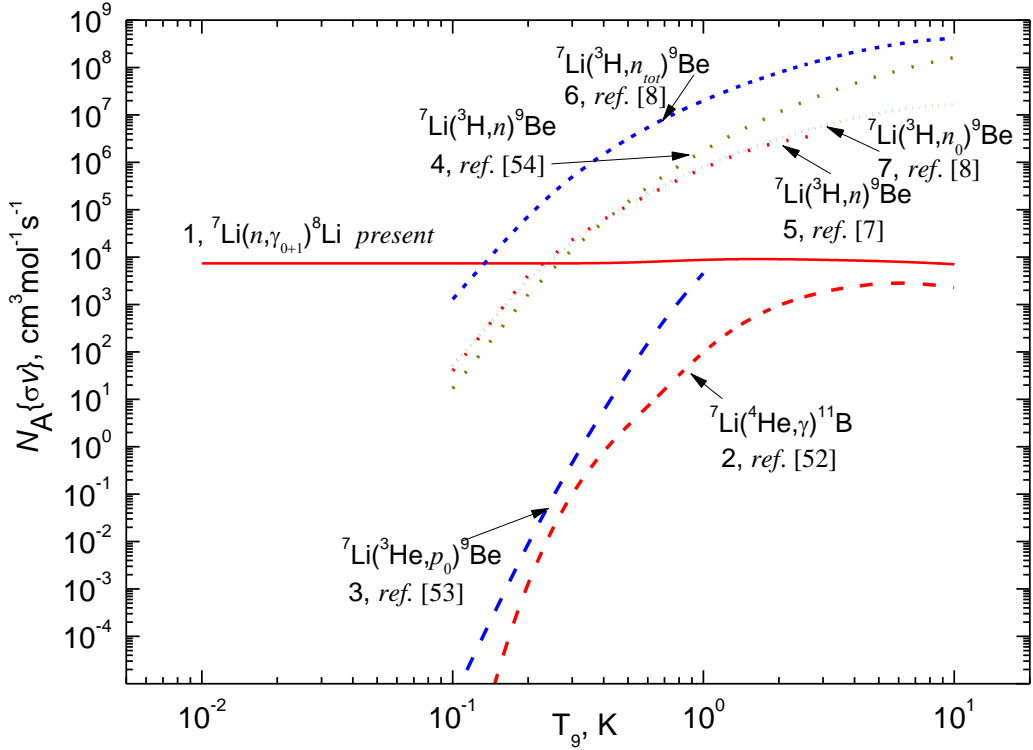

**Figure 5.** Comparison of the reaction rates of $^7$Li burning in exothermic reactions with n, $^3$H, $^3$He, and $^4$He. Curves are explained in the text.

It is reasonable to discuss the results of figure 5 along with known mass fraction (MF) shown in figure 6. These data have been adopted for our purposes and fit with [55].

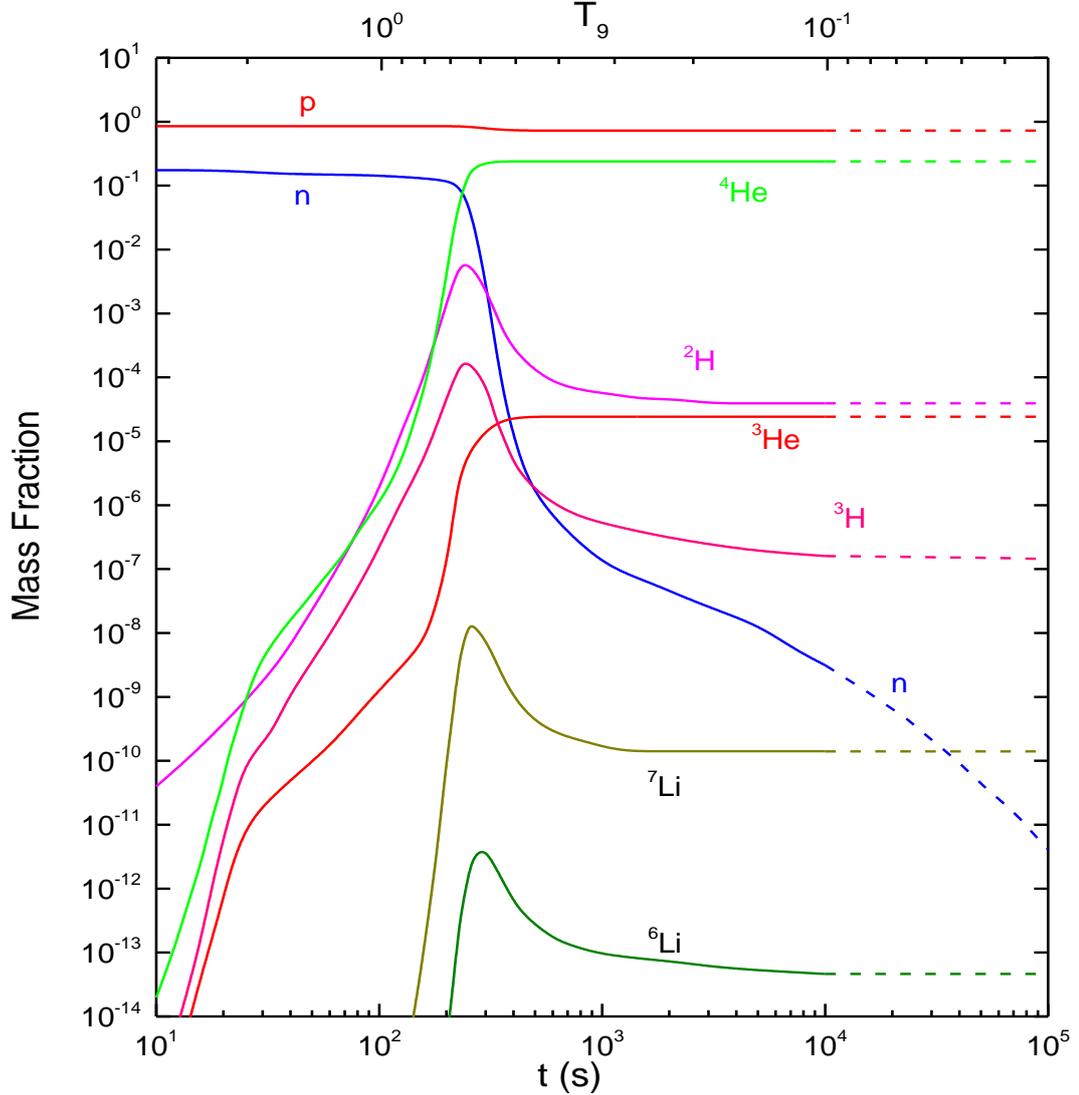

**Figure 6.** Mass fractions of light elements in BBN: solid curves [11]; dashed continuation – [56].

Our main goal declared in section 1 is to estimate the probability balance of different branching options (1.1)–(1.4) starting from the seed nucleus $^7$Li. While different chemical elements are already created, the possible burning of $^7$Li may be continued not only induced by neutrons, but triton (or $^3$He) and $^4$He nuclei also. Corresponding mass fractions of these species are seen in Fig 6.

Figure 5 illustrates the input of the corresponding reaction rates $\langle\sigma v\rangle$ relatively $^7$Li$(n,\gamma)^8$Li, $^7$Li$(^4$He, $\gamma)^{11}$B, $^7$Li$(^3$H,$n)^9$Be, and $^7$Li$(^3$He,$p_0)^9$Be. It can be seen from this comparison that the capture rate of $^7$Li$(^4$He,$\gamma)^{11}$B is located below our results of $^7$Li$(n,\gamma)^8$Li at any temperatures. The $\langle\sigma v\rangle$ of $^7$Li$(^3$He,$p_0)^9$Be behaves similarly. As concerns the $^7$Li$(^3$H,$n)^9$Be (curves 4–7), then starting roughly from $T_9 \sim 0.1$ corresponding value of $\langle\sigma v\rangle$ is laying higher than of neutron capture rate. Therefore, we suggest $T_9 = 0.1$ to be a benchmark in comparative analyses of RR$(x) = \langle\sigma v\rangle_x$ in units cm$^3$mol$^{-1}$s$^{-1}$ and MF$(x)$ for $x = n$, $^3$H, $^3$He, and, $^4$He:

$$\text{RR}(n) \approx 10^4 \quad \text{RR}(^3\text{H}) \approx 10^4 \quad \text{RR}(^3\text{He}) < 10^{-5} \quad \text{RR}(^4\text{He}) < 10^{-5} \quad (5.6)$$

$$\text{MF}(n) \approx 10^{-9} \quad \text{MF}(^3\text{H}) \approx 10^{-7} \quad \text{MF}(^3\text{He}) \approx 10^{-5} \quad \text{MF}(^4\text{He}) \approx 10^{-1}. \quad (5.7)$$

Firstly, let us remark that $^3$He may be excluded from the consideration. Triton branching (1.3) also cannot compete with chains (1.1), (1.2), and (1.4), i.e. *n* or $^4$He induced reactions if $T_9 < 0.1$. One can see that ratio MF(n) : MF($^4$He) $\approx 10^{-8}$ is compensated order of magnitude by the dominance of neutron capture reaction rate as RR(n) : RR($^4$He) > $10^9$. Thereby, we substantiated quantitatively the dominance of $^7$Li(*n*,γ)$^8$Li reaction yield comparing of the $^7$Li($^4$He,γ)$^{11}$B at the temperatures below 0.1–0.2 $T_9$.

## 6   Conclusion

Within the framework of the MPCM for the *n*$^7$Li system, we succeeded it is quite possible to correctly describe the experimental total cross sections for the reaction of neutron radiative capture on $^7$Li. It should be noted that on the basis of this model with the classification of orbital states according to Young diagrams, we have already considered almost 40 cluster systems see [13–18] (and refs therein). The early calculations [57–60] provide the investigations on the developing of nuclear chains both (1.1)–(1.4), and some others.

Based on the theoretical total cross sections obtained, the $^7$Li(*n*,γ)$^8$Li capture reaction rates were calculated and compared with the results of other studies and other reactions. A simple parametrization of the obtained reaction rate is proposed, which is useful for other works in this field and for applied problems.

It can be seen from the obtained results that, at all energies, the reaction rate of the neutron radiative capture exceeds the capture rate of alpha particle. However, both of these reaction rates at temperatures above 0.1–0.2 $T_9$ turn out to be much lower than the rate of the $^7$Li($^3$H,*n*)$^9$Be reaction proceeding due to the strong interactions. However, at lower temperatures, the $^7$Li($^3$H,*n*)$^9$Be reaction rate rapidly decreases and becomes less than the results we obtained for the neutron capture.

The $^7$Li($^3$He,*p*$_0$)$^9$Be reaction rate, due to a twofold increase of the Coulomb barrier, is lower than the reaction rate with tritium $^7$Li($^3$H,*n*)$^9$Be. Even at temperatures below 0.5 $T_9$, it plays practically no role in comparison with the neutron capture reaction.

Finally, we argued quantitatively that nuclear chains (1.1) and (1.2) are most probable at the temperatures less than 0.1–0.2 $T_9$ compering other branching, as the neutron capture rate is predominant.

**Acknowledgments**

This work was supported by the Grant of Ministry of Education and Science of the Republic of Kazakhstan through the program No. BR05236322 "Investigations of physical processes in extragalactic and galactic objects and their subsystems" under the theme "Studying thermonuclear processes in stars and primordial nucleosynthesis of the Universe" through the Fesenkov Astrophysical Institute of the National Center for Space Research and Technology of the Aerospace Committee of the Ministry of Digital

Development, Innovations and Aerospace Industry of the Republic of Kazakhstan.